\documentclass[sigconf]{acmart}

\usepackage{algorithm}
\usepackage{algpseudocode}

\usepackage{booktabs,graphicx,multirow}
\usepackage{url}
\AtBeginDocument{%
  }

\copyrightyear{2026}
\acmYear{2026}
\setcopyright{cc}
\setcctype{by-nc-nd}
\acmConference[CHI EA '26]{Extended Abstracts of the 2026 CHI Conference on Human Factors in Computing Systems}{April 13--17, 2026}{Barcelona, Spain}
\acmBooktitle{Extended Abstracts of the 2026 CHI Conference on Human Factors in Computing Systems (CHI EA '26), April 13--17, 2026, Barcelona, Spain}
\acmDOI{10.1145/3772363.3799327}
\acmISBN{979-8-4007-2281-3/2026/04}

\begin{document}

\title{DiverXplorer: Stock Image Exploration via Diversity Adjustment for Graphic Design}

\author{Antonio Tejero-de-Pablos}
\orcid{1234-5678-9012}
\affiliation{%
  \institution{CyberAgent}
  \city{Shibuya}
  \state{Tokyo}
  \country{Japan}}
\email{antonio_tejero@cyberagent.co.jp}

\author{Sichao Song}
\affiliation{%
  \institution{CyberAgent}
  \city{Shibuya}
  \state{Tokyo}
  \country{Japan}}
\email{song_sichao@cyberagent.co.jp}

\author{Naoto Ohsaka}
\affiliation{%
  \institution{CyberAgent}
  \city{Shibuya}
  \state{Tokyo}
  \country{Japan}}
\email{naoto.ohsaka@gmail.com}

\author{Mayu Otani}
\affiliation{%
  \institution{CyberAgent}
  \city{Shibuya}
  \state{Tokyo}
  \country{Japan}}
\email{otani_mayu@cyberagent.co.jp}

\author{Shin’ichi Satoh}
\affiliation{%
  \institution{CyberAgent}
  \city{Shibuya}
  \state{Tokyo}
  \country{Japan}}
\email{satoh@nii.ac.jp}

\renewcommand{\shortauthors}{Tejero-de-Pablos et al.}

\begin{abstract}
  Graphic designers explore large stock image collections during open-ended or early-stage design tasks, yet common tools emphasize relevance and similarity, limiting designers’ ability to overview the design space or discover visual patterns. We present an image exploration prototype that enables stepwise adjustment of diversity, allowing users to transition from diverse overviews to increasingly focused subsets during exploration. Our approach implements diversity control via determinantal point process (DPP)-based sampling and exposes diversity-similarity tradeoffs through interaction rather than static ranking. We report findings from a pilot study with professional graphic designers comparing our technique to baselines inspired by current tools in open-ended image selection tasks. Results suggest that stepwise diversity control supports early-stage sensemaking and comparison of visual patterns, while revealing important tradeoffs: diversity aids discovery and reduces backtracking, but becomes less desirable as exploration progresses. We aim to provide a novel perspective on how to implement transitions between diversity and similarity. Our code is available at \url{https://github.com/CyberAgentAILab/DiverXplorer}.
\end{abstract}


\begin{teaserfigure}
    \centering
    \includegraphics[width=0.9\textwidth]{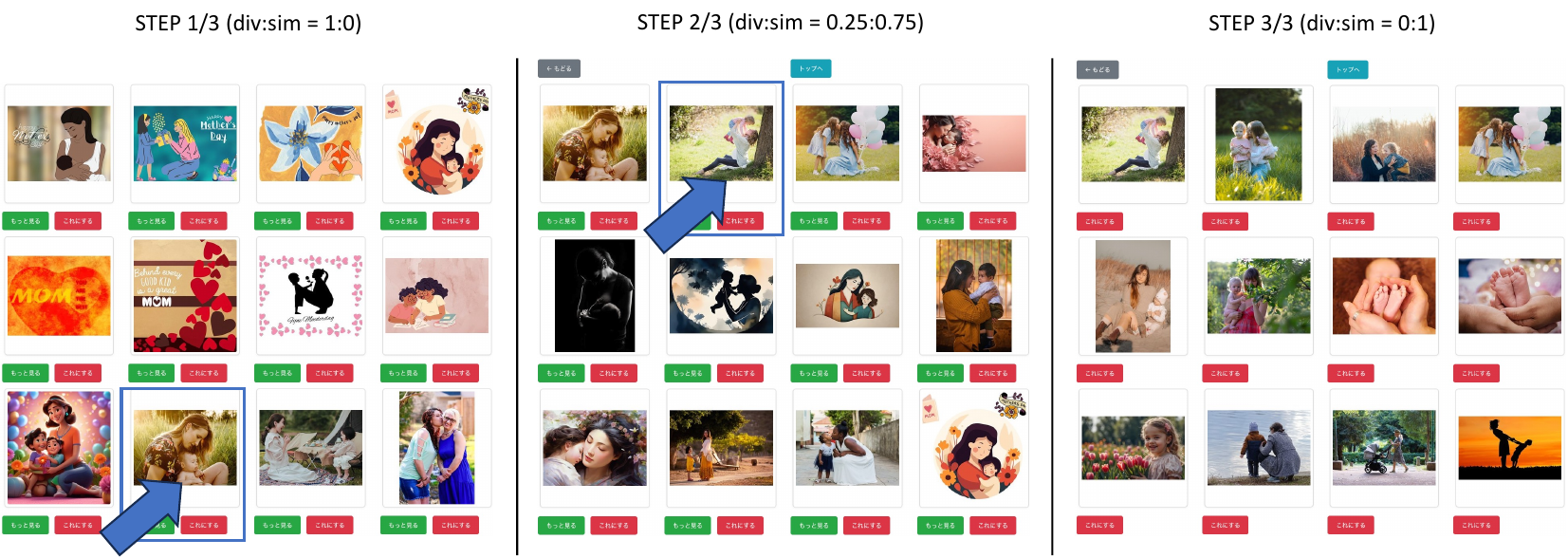}
    \caption{We divide image exploration into multiple steps (three in the image), for which the visualized images are optimized to meet a diversity:similarity ratio that gradually shifts from diverse to similar. The blue arrow represents the selected image on each step.}
    \Description{Overview of the proposed exploration method. By clicking an image on each step, the images displayed on the next step vary from diverse to similar.}
    \label{fig:diverxplorer}
\end{teaserfigure}


\maketitle

\section{Introduction}

Graphic designers rely on large-scale image databases---such as stock and AI-generated image collections \cite{kim2023effect}---when developing visual concepts for posters, advertisements, and other creative artifacts \cite{westerman2007supporting}. These tasks often begin with underspecified goals, requiring designers to explore and compare possible directions before committing to one \cite{white2009exploratory}.

Most existing image exploration tools prioritize relevance ranking and visual similarity \cite{chen2020wireframe,kang2021metamap,zhou2025productmeta}, typically presenting results in scroll-based grids (e.g., Adobe Stock) or similarity-driven navigation (e.g., Pinterest). While effective for targeted search \cite{barnaby2024photoscout}, these approaches can limit exposure to alternative styles and hinder designers’ ability to form an overview of the design space \cite{song2006diversifying}. Prior HCI research shows that such limitations can contribute to fixation and constrain exploratory sensemaking, particularly in ambiguous or open-ended tasks \cite{kim2022mixplorer}.

At the same time, diversity is not uniformly beneficial throughout the design process: highly diverse stimuli may support early exploration, whereas later stages often benefit from comparing closely related alternatives. This tension raises an open design question: how should image exploration interfaces support transitions between diversity and similarity, and how do such approaches compare to tools designers commonly use?

We explore this question through \textit{DiverXplorer}, an interaction technique that exposes diversity as a controllable dimension during image exploration (Fig.~\ref{fig:diverxplorer}). Rather than presenting a fixed ranking or similarity structure, DiverXplorer displays image subsets across multiple steps, each corresponding to a different point along a diversity-similarity spectrum. Users progress by making selections that gradually narrow the exploration space. The system was developed in collaboration with professional graphic designers to reflect limitations of existing tools and inform its functionality.

We evaluate DiverXplorer through a case study in which designers completed four open-ended stock image selection tasks using different exploration tools. We logged interaction behavior and conducted interviews to capture designers’ experiences. Our findings indicate a positive trend toward diversity-similarity-controlled exploration and suggest that this approach is more appropriate than common tools for open-ended graphic design tasks.

In summary, this work contributes: (1) a diversity-adjustable, stepwise sampling technique for stock image exploration (\textit{DiverXplorer}); (2) a comparative case study with twenty-two professional graphic designers; and (3) empirical insights into how image exploration tools can better support graphic design workflows.

We treat DiverXplorer as a design probe rather than a definitive solution, focusing on how stepwise diversity control shapes exploration behavior, user perceptions, and tradeoffs in stock image browsing. While diversity and stepwise interfaces have been studied in creative and ideation contexts, we argue that stock image exploration for graphic design represents a distinct and underexplored task setting.

\textit{Clarifications.}
This work does not address query-based image search; images are explored within a pre-existing collection or retrieved result set. We evaluate exploration strategies rather than propose a finished system. Here, “diversity” refers to visual differences among stock images, not demographic attributes.

\section{Background}
\label{sec:formative}

\paragraph{Formative study.}
To ground our work in designers’ actual practices, we interviewed four professional graphic designers working at a subsidiary company in Japan. They described a process in which, after receiving a design task, they first frame the problem and form an initial, broad idea, then explore stock images to translate this idea into a concrete visual direction. Selected images are then used to create a draft, which is validated before final delivery.

Designers emphasized that the exploration phase is particularly critical when design briefs are sparse and initial ideas remain vague. They described a need to first obtain an overview of possible visual patterns in the image collection and then gradually refine their ideas. However, they consistently characterized the commercial tools they use daily (e.g., Adobe Stock, Pixta, Pinterest) as a bottleneck, requiring repeated back-and-forth navigation through large image collections to identify suitable images. These tools primarily rely on relevance-based ranking, similarity-based exploration from a selected image, and scroll-based interfaces that present long, continuous result lists---interaction patterns that designers felt hindered both overview formation and iterative refinement.

\paragraph{Previous work.}
Prior studies show that designers often struggle to articulate precise queries early in the design process and instead rely on browsing, comparison, and reinterpretation of visual materials \cite{koch2020semanticcollage,bunian2021vins,son2024genquery}. Empirical work further indicates that preferred images tend to remain semantically relevant while being conceptually diverse enough to inspire new ideas \cite{westerman2007supporting}, particularly in early-stage or open-ended design tasks common in stock image exploration \cite{kim2022mixplorer}. This challenge is compounded by evidence that both retrieval- and generative-AI–based systems can encourage fixation when early outputs dominate subsequent exploration \cite{choi2025expandora,ko2023large,wadinambiarachchi2024effects,anderson2024homogenization}.
To mitigate fixation, recent design support systems emphasize visual distance as a driver of inspiration, motivating diversification strategies such as clustering and controlled sampling \cite{ranscombe2024inspiration,park2025leveraging,mozaffari2022ganspiration,liu2024sample,van2009visual,capannini2011efficient,radlinski2006improving,son2024genquery}. Diversification has been explored in both image-based design tools \cite{guo2021vinci,kim2022mixplorer,han2025poet} and prompt-based interaction with generative models \cite{almeda2024prompting}. In image search, query expansion and mixed-initiative approaches support evolving user intent and serendipitous discovery \cite{ye2024contemporary,koch2020imagesense,kovacs2018context}. However, prior work also shows that excessive or poorly structured diversity can overwhelm users or introduce irrelevant results \cite{fu2013meaning,cai2023designaid}.

Most existing systems embed diversity implicitly and provide limited user control, constraining designers’ ability to balance relevance, similarity, and diversity. While some approaches incorporate diversity directly into retrieval or generation models, jointly optimizing relevance and diversity remains challenging, leading many systems to introduce diversity as a post-processing step during exploration \cite{baggio2023multi}. Clustering-based methods commonly support overview-first exploration \cite{johnson1967hierarchical,shneiderman2003eyes}, but are sensitive to data distribution, offer limited diversity control, and do not support smooth refinement \cite{carpineto2012evaluating,oesterling2024multi}. In contrast, diversity-optimizing methods such as determinantal point processes (DPPs) enable explicit control over the relevance-diversity tradeoff \cite{macchi1975coincidence,kulesza2012determinantal}.

Beyond diversification, HCI research highlights the importance of coarse-to-fine, abstract-to-specific refinement in exploratory tasks \cite{smith2025fuzzy}. Iterative refinement supports ideation across domains, including creativity support tools \cite{wang2022interpretable}, color design systems \cite{shi2024exploring}, and interfaces for refining prompts or sketches in generative image creation \cite{tao2025designweaver,lin2025inkspire}. Similar principles have been applied to image exploration through grid-based interfaces that enable gradual narrowing of design alternatives \cite{mohiuddin2020interactive,koyama2020sequential}. Together, these findings motivate the development of an approach that combines explicit diversity optimization with stepwise refinement for stock image exploration.

\paragraph{Implementation decisions of DiverXplorer.}
Informed by feedback from professional graphic designers and prior human–computer interaction research, we identified three core functionalities for DiverXplorer: (1) incorporating image diversity when selecting which images to display, (2) varying the level of diversity as exploration progresses, and (3) supporting stepwise user interaction to enable coarse-to-fine refinement.

\paragraph{Research question.}
Rather than focusing on prompt-based image generation or query refinement, we examine stock image exploration as a distinct design task and treat diversity as an interactive dimension that can be adjusted throughout the exploration process. Accordingly, we investigate the following \textbf{research question}:
How should stock image exploration interfaces support transitions between diversity and similarity, and how does a stepwise diversity-controlled approach compare to baseline implementations inspired by tools commonly used by designers?

\section{\textit{DiverXplorer}: stepwise diversity control}
\label{sec:diverxplorer}

\textit{DiverXplorer} (Fig.~\ref{fig:diverxplorer}) is an image exploration technique that supports stepwise refinement by progressively adjusting the diversity of displayed images. At each exploration step, the system presents a fixed-size subset of images sampled from a larger collection, where early steps emphasize diversity to provide an overview of visual patterns, and later steps increasingly emphasize similarity to support focused comparison and selection. Conceptually, diversity and similarity are treated as opposite ends of the same axis, enabling a gradual overview-to-refine transition.

To operationalize diversity, we employ a diversity-optimizing sampling strategy based on determinantal point processes (DPPs) \cite{macchi1975coincidence,kulesza2012determinantal}, which have been widely used for selecting diverse subsets from large collections.
Traditional DPPs sample subsets that are maximally spread out in an image embedding space, favoring sampling of dissimilar images. By dynamically constraining the similarity distance of the images sampled, our technique enables explicit control over the tradeoff between diversity and relevance. In our setting, we extract embeddings via the CLIP image encoder~\cite{radford2021learning}, as its feature representations balance better both visual and semantic relationships.

Exploration proceeds iteratively: users select an image at each step, which conditions the next subset of images sampled by the system. Diversity constraints are gradually relaxed according to a predefined schedule, resulting in increasingly similar image sets as exploration progresses. To support visual comparison, sampled images are displayed in a grid layout, which allows users to compare multiple candidates side-by-side and aligns with common image browsing practices \cite{koyama2020sequential}. Each step displays the same number of images, enabling consistent comparison across refinement stages.

Through this combination of explicit diversity optimization and stepwise interaction, DiverXplorer supports coarse-to-fine image exploration while treating diversity as an interactive dimension rather than a fixed hyperparameter. The appendix contains detailed algorithms and comparisons, implementation details, parameter settings, and pseudocode.

\section{Exploratory user study}

We conducted a user study (August 2025, approved by the ethics committee at our organization) to examine how stepwise diversity control shapes designers’ image exploration experiences in open-ended stock image selection tasks, and to evaluate \textit{DiverXplorer} against commonly-used baselines.

\paragraph{Participants.}
Twenty-two professional graphic designers collaborated with this research: four participated in the formative study (Sec.~\ref{sec:formative}), two in a rehearsal session to refine materials and procedure, and sixteen participated in the main study. Participants were employed full- or part-time at a subsidiary company in Japan; they included twelve women and four men, spanning their 20s$\sim$40s with 2$\sim$15 years of professional experience.

\paragraph{Tasks and dataset.}
During the formative study, we defined four open-ended design briefs that intentionally provide sparse constraints (topic, target audience, and intended feeling/message), reflecting scenarios where designers must explore before committing to a direction.
\begin{itemize}
    \item Task A: A design for a lipstick make-up aimed at middle-aged women with the message of ``you will definitely look beautiful with this product'' (query:~\textit{lipstick woman beautiful}).
    \item Task B: A design to celebrate Mother's Day aimed at young mothers with the feeling of ``spending happy times with their children'' (query:~\textit{mother's day child}).
    \item Task C: A design for a winter vacation plan aimed at couples in their 30s that conveys a ``romantic atmosphere'' (query:~\textit{winter romantic couple}).
    \item Task D: A design for a birthday party aimed at elementary school children with an ``it will be so fun'' message (query:~\textit{birthday party kids}).
\end{itemize}
Each task was paired with a fixed set of 600 stock images (downloaded from Pixabay\footnote{\url{https://pixabay.com} (last accessed 2025-09-11)}) collected via the corresponding query; to avoid confounds from query formulation, participants did not enter queries and instead explored within the provided result set.
The appendix includes example images for each task.

\begin{figure*}[t]
    \centering
    \includegraphics[width=0.9\textwidth]{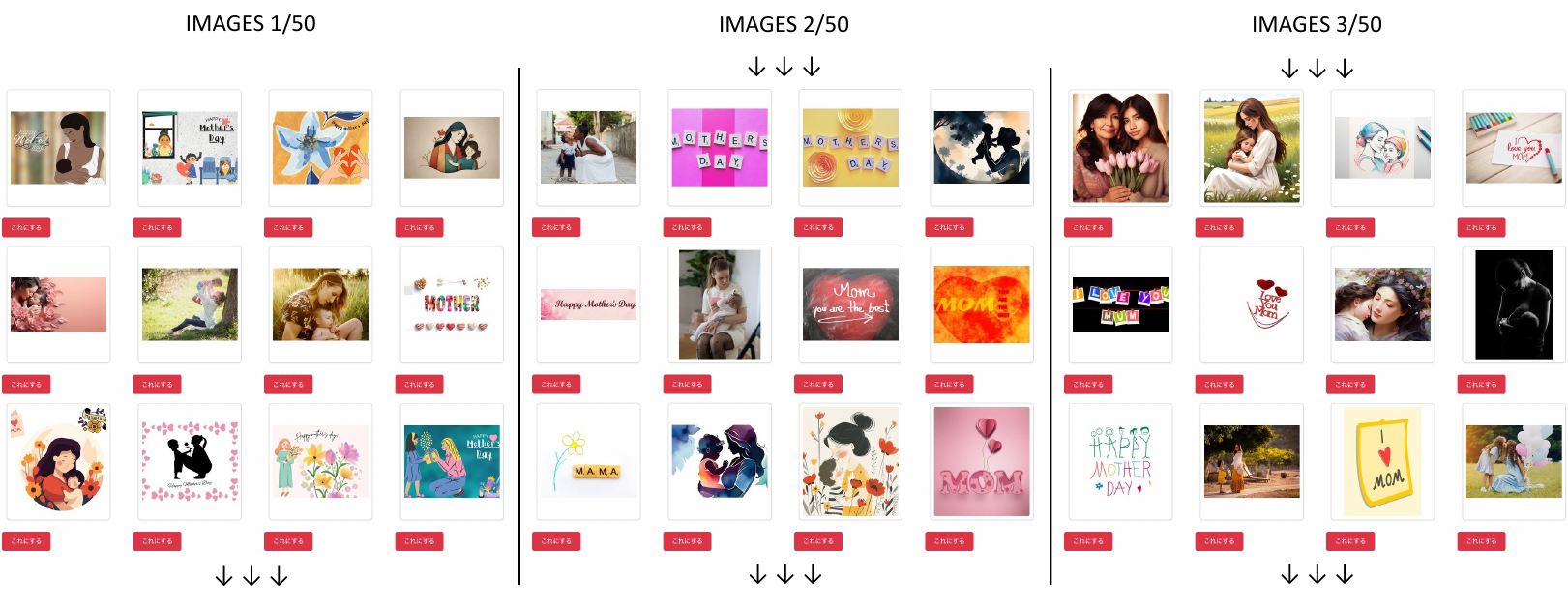}
    \caption{\textit{Scroll} displays all images on a single screen, sorted by relevance to the task (i.e., ``A design to celebrate Mother's Day aimed at young mothers with the feeling of "spending happy times with their children'').}
    \Description{Overview of the Scroll tool. Unlike stepwise methods, images are listed on a single screen, and users need to scroll-down to explore them.}
    \label{fig:scroll}
\end{figure*}

\begin{figure*}[t]
    \centering
    \includegraphics[width=0.9\textwidth]{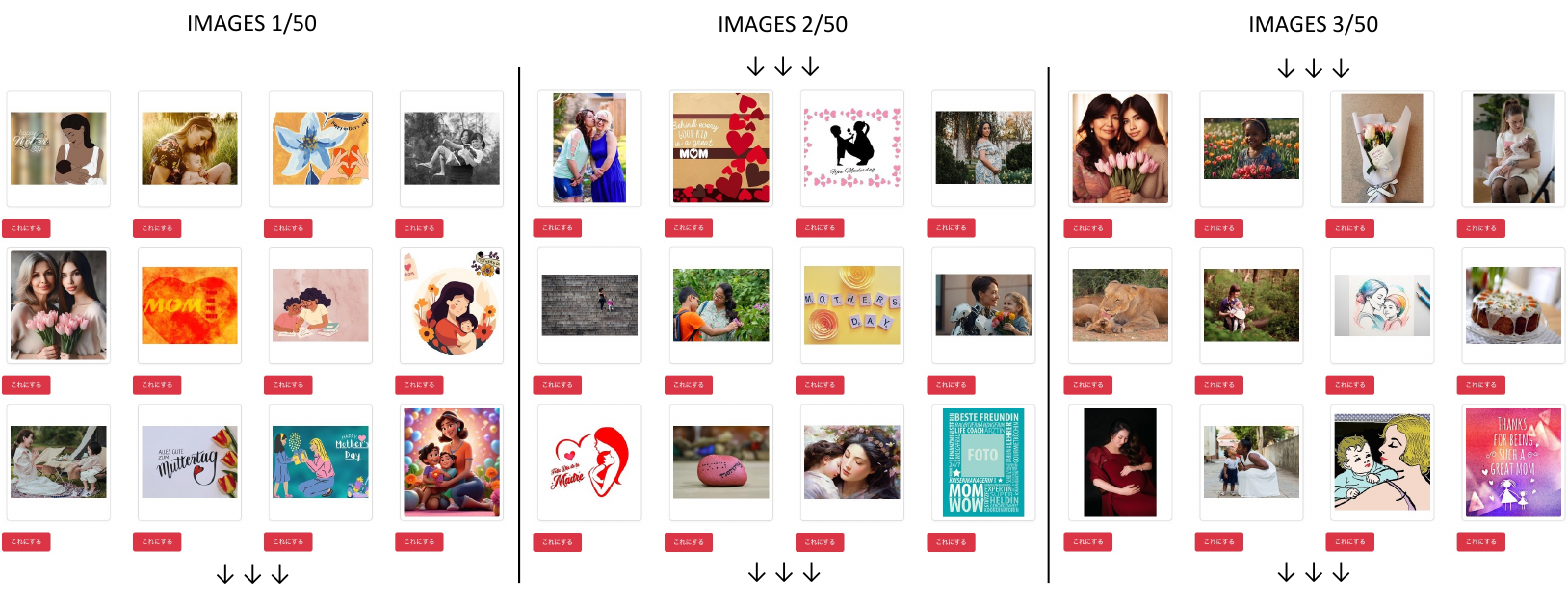}
    \caption{\textit{Scroll+div} displays all images on a single screen, and uses the DPP algorithm to rerank them by diversity and relevance to the task (i.e., ``A design to celebrate Mother's Day aimed at young mothers with the feeling of "spending happy times with their children'').}
    \Description{Overview of the Scroll+div tool. Unlike stepwise methods, images are listed on a single screen, and users need to scroll-down to explore them.}
    \label{fig:scroll+div}
\end{figure*}

\begin{figure*}[t]
    \centering
    \includegraphics[width=0.9\textwidth]{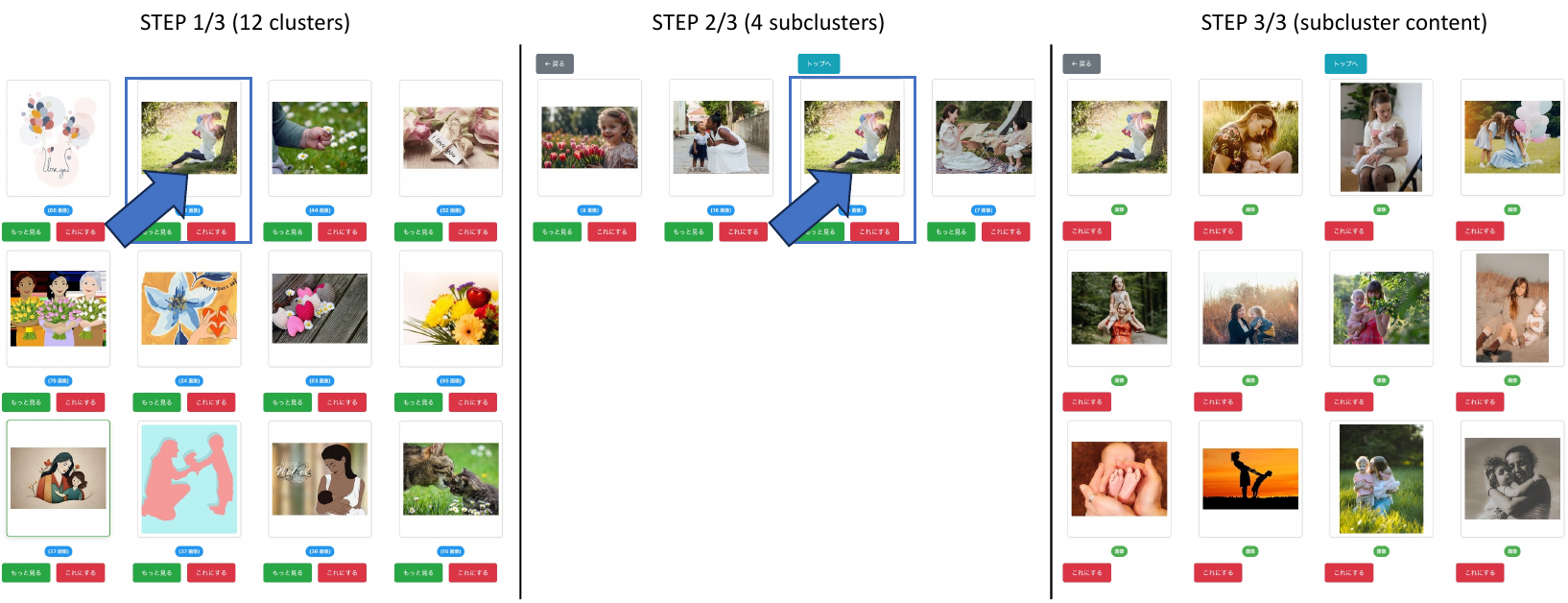}
    \caption{\textit{Clustering} generates a set of subclusters/steps (e.g., 3 in the image) for which only similar images are linked to. The blue arrow represents the selected image on each step. The hierarchy is data-dependent and a constant number of subclusters cannot be guaranteed (e.g., only four options in step 2).}
    \Description{Overview of the Clustering tool. By clicking an image on each step, only similar images to the one selected are displayed.}
    \label{fig:clustering}
\end{figure*}

\paragraph{Conditions: baselines and ablation.}
To ensure realistic comparisons, and with the guidance provided in the formative study, we implemented four minimalist tools with a shared visual style.
\begin{enumerate}
    \item \textbf{Scroll} (relevance-ranked grid with scrolling, Fig.~\ref{fig:scroll}): a baseline approximating commercial stock search interfaces such as Adobe Stock and Pixta. Users can scroll up and down to explore the collection, and select the desired image by clicking the red button below each image.
    
    \item \textbf{Scroll+div} (diversified re-ranking with scrolling, Fig.~\ref{fig:scroll+div}): identical to Scroll but re-ordered using a diversity-aware ranking (i.e., DPP with max. diversity), testing whether ``naive'' diversification without stepwise interaction is sufficient.

    \item \textbf{Clustering} (stepwise similarity refinement, Fig.~\ref{fig:clustering}): highly relevant practically---as indicated by designers regarding Pinterest's similarity-based exploration---and algorithmically---as using clustering centroids for diversity is widely-used for idea refinement exploration~\cite{shneiderman2003eyes,guo2021vinci,mozaffari2022ganspiration,huang2023diversity,matsui2025lotusfilter}. Users can choose from an initial set of clusters and, by pressing the green ``See more'' button, they can visualize clusters of similar images, until the last step where the images of the selected final cluster are displayed. The red ``Choose'' button allows selecting the desired image at any exploration step. Users can go back to previous steps via a black ``Back'' and a blue ``Top'' buttons and choose different images. The purpose of this baseline is to evaluate an alternative stepwise method to our \textit{DiverXplorer}, in which images can be selected in several steps that narrow down the target. However, in clustering, diversity is a side-effect of grouping similar images together, and is not explicitly optimized as in our proposed method. For example, when selecting a real photo with a mother and a child in a park, illustrations and photos without kids are not further displayed, and only similar images are proposed.

    \item \textbf{DiverXplorer} (stepwise diversity control, Fig.~\ref{fig:diverxplorer}): our proposed technique combining stepwise interaction with explicit diversity optimization. It also features a green ``See more'', red ``Choose'', black ``Back'' and blue ``Top'' buttons for navigation. However, unlike \textit{Clustering}, \textit{DiverXplorer} displays images that are not necessarily similar to the one selected, if the trade-off diversity/relevance is met. For example, when selecting a photo with a mother and a child in a park, images without kids are not displayed in the next step, but illustrations of a mother and a child and photos that are not in a park are also proposed. The purpose of this tool is evaluating if diversity-based exploration featuring stepwise iterative refinement is suitable for designers' work.
\end{enumerate}


\paragraph{Procedure.}
Study sessions were conducted one-on-one over an online meeting platform. After an introduction, participants completed four trials, each consisting of one task using one tool (counterbalanced via a Graeco-Latin square so that each participant used each tool exactly once and task--tool pairings were balanced across participants; details in appendix). Before each trial, participants completed a brief warm-up to learn the tool. Participants then explored the image set and selected the image they considered most suitable for the brief; there was no time limit. After each task, they answered a questionnaire to evaluate the effectiveness, usability and general satisfaction of the tool on a 7-point Likert scale. After all four tasks, participants provided a comparative ranking of the four tools and participated in a semi-structured interview focused on (i) reasons for their ranking and (ii) impressions of diversity-based exploration. Interviews were transcribed for analysis.

\paragraph{Measures and analysis.}
We collected (1) interaction logs, (2) answers to the questionnaire, and (3) interview responses. Quantitative analyses focused on descriptive comparisons across conditions (questionnaire and ranking scores), while qualitative analysis summarized recurring themes in participants’ reflections, using representative excerpts to ground claims (specific implementations, questionnaire and other details in appendix). We treat findings as exploratory observations rather than confirmatory evidence.
 
\section{Results and analysis}

\begin{figure}[t]
    \centering
    \includegraphics[width=0.9\linewidth]{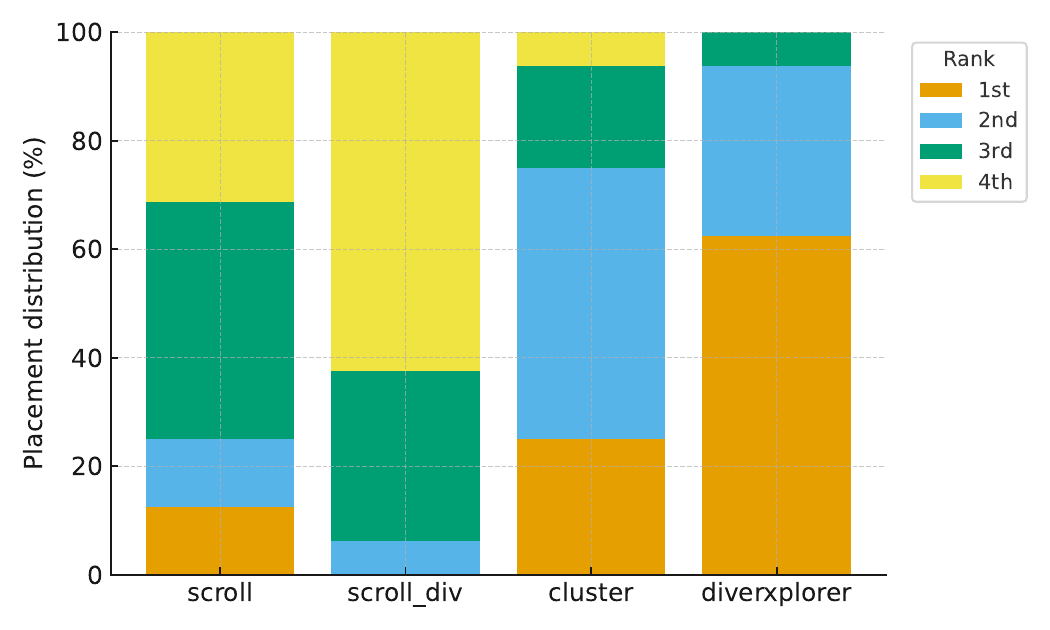}
    \caption{Ranking results. \textit{DiverXplorer} was ranked 1st by more than 60\% of the designers.}
    \Description{Bar graph with colors indicating how many times each tool was ranked 1st (orange), 2nd (blue), 3rd (green), 4th (yellow).}
    \label{fig:ranking}
\end{figure}

In the questionnaire, as a general pattern, the stepwise exploration tools obtained higher scores than scroll-based ones (full analysis in the appendix).
Fig.~\ref{fig:ranking} depicts the ranking results of the interview comparing all tools, which match the trend observed in the questionnaire. Our proposed \textit{DiverXplorer} was ranked first more than 60\% of the times, becoming the preferred tool by graphic designers.
Next, we analyze the interview responses, where the transcriptions of the interviews became insights through Reflexive Thematic Analysis. Participant counts are reported as (n/16).

\paragraph{Stepwise refinement supported sensemaking but introduced interaction costs.}
Designers reported that scroll-based tools exposed too many options at once, making decisions difficult (8/16). In contrast, stepwise tools enabled them to incrementally remove undesired patterns and clarify intent (9/16), with some valuing the ability to pause and reflect at each step (2/16). Such an increase in usefulness also comes with a slight interaction friction; some designers mentioned that stepwise refinement introduced repeated button clicks (4/16) and reduced predictability compared to scroll-based interfaces (4/16).

\paragraph{Diversity supported discovery when structured, but hindered refinement when unbounded.}
Designers criticized \textit{Scroll} for making it difficult to encounter new patterns (4/16). Although \textit{Scroll+div} was not generally preferred, many acknowledged that it surfaced unexpected styles they would not have considered otherwise (13/16). Diversity was perceived as more usable in \textit{DiverXplorer} than in \textit{Scroll+div} (8/16), as stepwise refinement presented a controlled number of images per step while enabling cross-style comparison (5/16). In contrast, designers felt that in \textit{Clustering} diversity faded too quickly, as subclusters became homogeneous right after the first step (5/16) and uneven cluster sizes limited choice mid-exploration (4/16), often forcing backtracking; this aligns with prior findings on clustering-based diversification~\cite{guo2021vinci}.

\paragraph{Designers’ preferences shifted as exploration converged.}
While diversity was valued during early exploration, designers reported that its usefulness decreased once they committed to a concrete style. In \textit{Scroll+div}, unrelated alternatives were seen as obstructive (6/16), and designers preferred browsing similar images for fine-grained comparison. Accordingly, grouping similar images remained important for convergence: designers favored \textit{Scroll} over \textit{Scroll+div} due to clearer similarity structure (12/16), and some found \textit{Clustering} advantageous for detailed comparison because of its predictable, folder-like organization (4/16). 
Created according to the designers' specifications, \textit{DiverXplorer} not only reduces diversity as exploration progresses, it does it exponentially so early stages dispose unrelated images and later stages allow for precise refinement.

\section{Lessons Learned}

The following lessons link to our research question by clarifying how stock image exploration interfaces should support transitions between diversity and similarity, and distinguishing our approach from commonly used baselines.

\paragraph{Diversity is a phase-dependent resource that must be transitioned over time.}
Diversity was most valuable during early exploration, helping designers orient themselves within the design space and discover alternative styles, but became less desirable as designers converged on a specific direction. Treating diversity as uniformly beneficial led to confusion during refinement, while excessive similarity too early risked fixation. Supporting explicit transitions between diversity and similarity is therefore essential.

\paragraph{Stepwise interaction enables interpretable diversity–similarity transitions.}
Diversity-only approaches surfaced novel content but were difficult to interpret without intermediate structure, while clustering-based approaches collapsed diversity too quickly. Stepwise interaction allowed designers to reason about alternatives incrementally, making diversity actionable rather than overwhelming. Compared to the other baselines, stepwise diversity control better supported gradual refinement without forcing premature commitment.

\paragraph{Exploration tools should support divergence-convergence shifts.}
Designers needed to fluidly move between exploratory (divergent) and comparative (convergent) modes. Baseline tools tended to enforce a single mode---either broad diversity or strict similarity---resulting in mismatches with real workflows. In contrast, stepwise diversity control supported smoother transitions between modes, aligning closely with how designers explore, refine, and commit to visual directions.

\section{Limitations and future work}

This exploratory study focused on designers’ perceptions and experiences in controlled stock image selection tasks and does not capture downstream design activities such as layout composition or integration with text and branding. The participant pool was limited in size and demographics, and the evaluation relied primarily on subjective and comparative measures rather than long-term or in-the-wild use. Future work should examine behavioral traces of exploration in greater depth, involve larger and more diverse populations, and study how diversity control interacts with later stages of the design pipeline. In addition, given the computational cost of DPP algorithm's sequential sampling, scaling diversity-aware exploration to larger image collections and incorporating alternative feature representations remain important directions for further investigation.









\bibliographystyle{ACM-Reference-Format}
\bibliography{main}
\newpage

\appendix

\section{Algorithmic and implementation details of DiverXplorer}

Determinantal Point Processes (DPPs) have been used in machine learning for optimal diverse subset selection. This diversity can be measured by calculating distances between samples in the image feature space (i.e., sample correlation in DPP).
While alternative diversity metrics exist (e.g., Vendi Score \cite{friedman2023vendi}, Magnitude \cite{limbeck2024metric}), empirically no significant difference was found betweeen DPP and Vendi in our scenario, while Magnitude performed worse.

\begin{figure}[ht]
    \centering
    \includegraphics[width=0.9\linewidth]{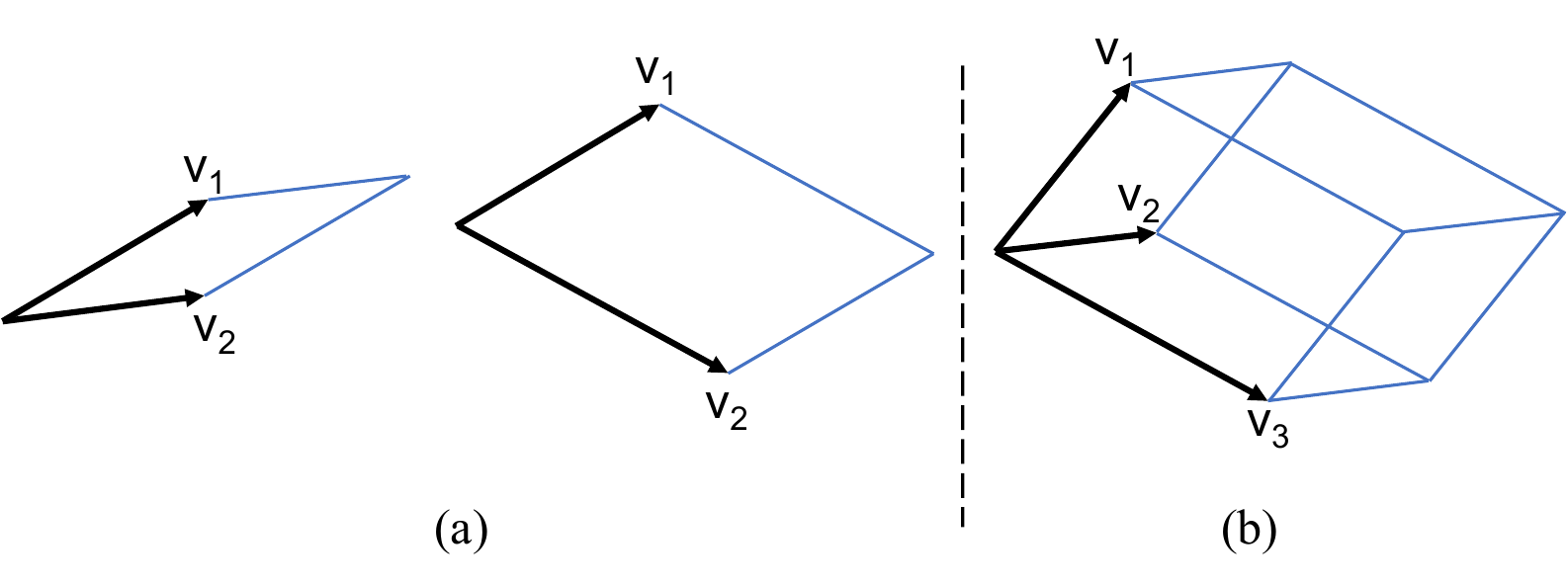}
    \caption{The determinant maximization of DPP is equivalent to the maximization of the volume comprised by the vectors in the feature space. (a) The further from each other samples are in the feature space, the larger the vector area. (b) Case of more than two dimensions.}
    \Description{Graphic visualization of the DPP algorithm in a 2D and 3D space. The larger the volume contained within the vectors, the more diversity the set has.}
    \label{fig:dpp}
\end{figure}

In a subset selection via DPP, the inclusion of one item makes the inclusion of other items less likely. The strengths of these negative correlations are derived from a kernel matrix that defines a global measure of similarity between pairs of items, so that more similar items are less likely to co-occur.
Formally, a DPP is a probability distribution over subsets of a finite ground set$~\mathcal{X}$ of $N$ elements, defined through a real, symmetric, positive semidefinite kernel matrix $K \in \mathbb{R}^{N \times N}$.
The probability of observing a subset ($X \subseteq \mathcal{X}$) is defined as
$\mathbb{P}(X) \propto \det{K_X}$,
where $\det{K_X}$ is the determinant of the similarity matrix restricted to the $M$ elements in $X$. 
The likelihood function has a geometric interpretation, as the square of the volume spanned by the elements of $X$ in an implicit feature space (Fig.~\ref{fig:dpp}).

Let us apply this theory to the diverse image exploration scenario.
Suppose we have feature vectors of the entire image set $V=\{v_1, ..., v_N\}, v_i \in \mathbb{R}^d$ and their corresponding relevance scores with respect to a certain topic $R=\{r_1, ..., r_N\}, r_i \in \mathbb{R}^1$. We can build a similarity matrix (also called Gram matrix) $L$ with entries $L_{ij} = v_i^\top v_j$ by using vectors sampled from $V$.
An optimal subset $S$ containing $M$ diverse samples is calculated via greedy sampling from $V$.
$S$ is calculated such that the determinant $\det L_S$ is maximized\footnote{This requires that $d> y$ for the determinant not to be $0$, condition largely met in our setting.}. The greedy approach builds $S$ iteratively; for a given set of feature vectors $V$, their relevance scores $R$, and the size of the diverse set $k$ we want to create (Alg.~\ref{alg:dpp}).

\begin{algorithm}[t]
\caption{Greedy DPP algorithm}
\label{alg:dpp}
\begin{algorithmic}[1]
    \Procedure{CreateOptimalDiversitySet}{$V$, $R$, $k$, $\omega_r$, $\omega_d$}
        \State Let $ S = \emptyset $
        \State Set $S_1 = V_1$
        \State Set $ V= V \setminus \{1\} $
        \State Calculate $\textrm{threshold}$
        \For{$ k = 2, \dots, K $}:
            \For{each $ i \in V $}:
                \State $\textrm{div}_i = \log \det{L_{S_{k-1} \cup i}}$
                \State $\Delta_i = \omega_r(\sum_{j \in S_{k-1}}r_j + r_i) + \omega_d \textrm{div}_i$
            \EndFor
            \State Choose $ i^* = \arg\max_{i \notin S_{k-1}} \Delta_i $,
            where $\textrm{div}_i < \textrm{threshold}$
            \State Set $ S_k = S_{k-1} \cup {i^*} $
            \State Set $ V = V \setminus \{i^*\} $
        \EndFor
        \State \textbf{return} $S$
    \EndProcedure
\end{algorithmic}
\end{algorithm}

Weights $\omega_r$ and $\omega_d$ can be set to encourage relevance over diversity and vice versa.
The relevance of the samples (i.e., $r_i \in R=0,\forall i$) can be calculated as a similarity score with the given exploration task (see Sec.~\ref{sec:implementation}).
The $\textrm{threshold}$ term is calculated by considering the maximum diversity and the exploration step; as diversity gradually decreases, we set the threshold as the quantile $q$ of the sorted values of diversity with respect to the sample selected as a root on that exploration step (i.e., $V_1$).

\begin{algorithm}[t]
\caption{Threshold calculation on each step}
\label{alg:threshold}
\begin{algorithmic}[1]
    \Procedure{CalculatingThreshold}{$V$, $q_e$}
        \State Let $ \textrm{div} = \emptyset $
        \State Let $ \textrm{root} = V_1 $
        \State Set $ V= V \setminus \{1\} $
        \For{each $ i \in V $}:
            \State $\textrm{div}_i = \log \det{L_{\textrm{root} \cup i}}$
        \EndFor
        \State Sort $\textrm{div}$ decreasingly
        \State Set $\textrm{threshold} = \textrm{quantile}(\textrm{div}, q_e)$
    \EndProcedure
\end{algorithmic}
\end{algorithm}

The value $q \in [0,1]$ can follow a linear $q_1=1, q_2=0.9, q_3=0.8, ..., q_{E-1}=0.2, q_E=0.1$ or an exponential decay $q_1=1, q_2=0.65, q_3=0.45, ..., q_{E-1}=0.1, q_E=0.05$. In this work, we opt for an exponential decay, as it fits better the narrowing down and final fine-selection of the creative process of designers. Additionally, we reranked the $k$ sampled images so similar samples appear together---an additional run of DPP minimizing diversity instead of maximizing it.

Our code is available at \url{https://github.com/CyberAgentAILab/DiverXplorer}.

\section{Experimental design details}

\subsection{Preliminary study}

To determine the most appropriate settings of our implementations we asked four graphic designers---the four participants of the formative study---to use all four tools and provide advice on specific implementation details. They were Japanese professional graphic designers at a subsidiary company in Japan. There were three women and one men, in their 20s and 30s, and their years of experience ranged from 5 to 15.

\paragraph{\textit{DiverXplorer} hyperparameters.} Following the designers' advice, we determined the most appropriate value for the following hyperparameters: grid size $k=4\times 4$ of the interface, number of steps $E=4$ to increase/decrease diversity, diversity decreasing quantile $q=[1, 7\cdot 10^{-2}, 5\cdot 10^{-3}, 3\cdot 10^{-4}]$, weights $\omega_r=12$ and $\omega_d=1$. Similarly, for the sake of comparison with the \textit{Clustering} tool, we generated $\approx16$ clusters for the top cluster layer (i.e., first exploration step) and a maximum hierarchy depth of $4$.

\paragraph{Image encoder} DPP calculates diversity in the feature space, and thus it is dependent on the image encoder used for feature extraction. We considered three state-of-the-art encoders: CLIP~\cite{radford2021learning}, DINOv2~\cite{oquabdinov2}, and ConvNeXt~\cite{liu2022convnet}. Among those, all four designers agreed that diversity calculated with CLIP features felt the most appropriate, as representations seemed to balance both visual and semantic relationships.

\begin{figure*}[t]
    \centering
    \includegraphics[width=\linewidth]{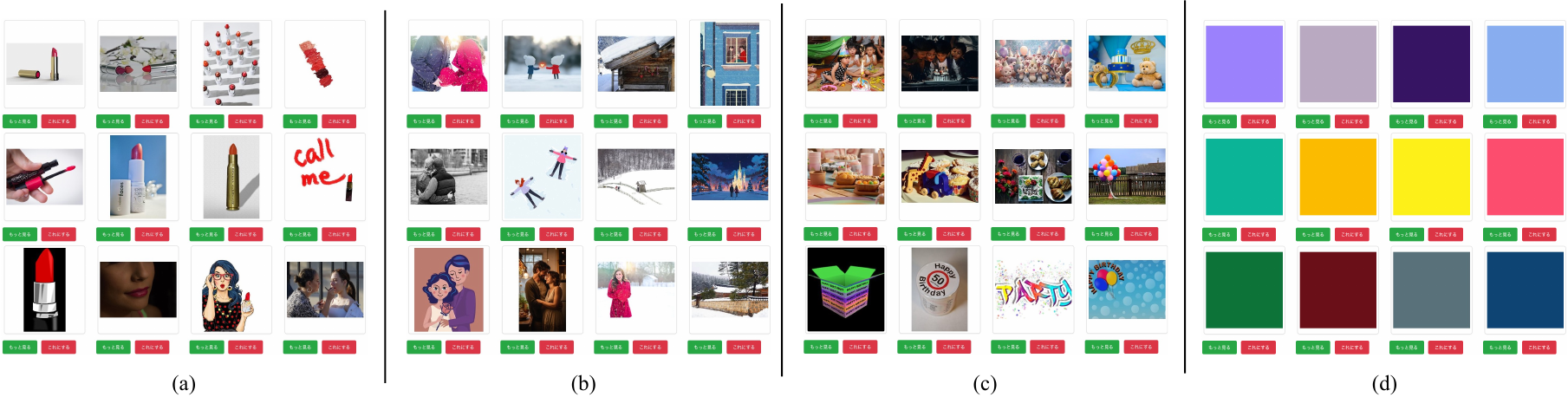}
    \caption{Sample images for (a) Task A: Lipstick makeup for middle-aged women, (b) Task C: Romantic winter for couples, (c) Task D: Birthday party for kids, (d) Warm-up task: Colors that convey coolness.}
    \Description{Sample images of each task except for Task B: Mother's Day: (a) Task A: Lipstick makeup for middle-aged women, (b) Task C: Romantic winter for couples, (c) Task D: Birthday party for kids, (d) Warm-up task: Colors that convey coolness.}
    \label{fig:tasks}
\end{figure*}

\paragraph{Task definition} We defined the exploration tasks for our user study during the preliminary study. While designers indicated that normally design requests are more specific than those used in the user study, they also acknowledged that these are the most likely to create a bottleneck in their regular job. Once a task is solved, it cannot be used to evaluate the next tool, since participants will already have decided on a specific stock image, and design support will probably not be effective.
Thus, we decided on a total number of four evaluation tasks---one per tool---, plus a warm-up task. Figure~\ref{fig:tasks} displays some sample stock images in the \textit{DiverXplorer} tool (see Fig.~\ref{fig:diverxplorer} for the Mother's Day task). Image (d) corresponds to the simple warm-up task used to explain the functioning of each tool. In this task, we asked designers to use the tool's scroll and buttons to try to find the color that better conveys ``coolness'' to them. In this process, designers got used to the tool before tackling the actual task in the user study.

\subsection{Implementation details}
\label{sec:implementation}

We implemented the \textit{Clustering} exploration tool via agglomerative clustering (Appendix~\ref{app:clustering} contains the details).
When the number of samples is large enough, hierarchical clustering allows distributing the samples among multiple layers of clusters, where the granularity becomes finer with each layer.

Relevance scores were calculated as the cosine similarity between the text of the task definition and the CLIP image features, namely, the CLIP score.

We implemented all methods in Python. We used \texttt{openai}/\\ \texttt{clip-vit-base-patch32} in HuggingFace transformers library for image feature encoding, and the \texttt{pywebio} library for the interfaces, which allows running the program on the browser via a local URL. In combination with the \texttt{ngrok} software\footnote{\url{https://ngrok.com} (last accessed 2025-09-11)}, our participants were able to remotely access our locally deployed tools from their browser in a one-on-one manner.

\subsubsection{Manipulation check}

We thoroughly debugged and tested our proposed system \textit{DiverXplorer} before the user study to check that it worked as expected. As a sanity check, we outputted the actual diversity values corresponding to the subset of stock images sampled on each step in Figure~\ref{fig:diverxplorer}: [-54.4947, -60.7311, -70.4490] for step 1, 2 and 3 respectively.
As expected, values shrink as the stepwise refinement progresses. Also, values are negative since the diversity metric ranges between [0,1] and its logarithm is taken (Algorithm~\ref{alg:dpp}).

\subsection{Clustering hyperparameters}
\label{app:clustering}

For the sake of reproducibility, we summarize here the hyperparameters used in the hierarchical clustering algorithm used to implement our \textit{Clustering} tool. Since using fixed parameters resulted in a performance too weak for comparison with our tool, parameters are automatically tuned based on the characteristics of the image data.

Key Features:
\begin{itemize}
    \item Adaptive cluster count: The algorithm dynamically determines the optimal number of clusters using silhouette score and Calinski-Harabasz index
    \item Data-driven early stopping: Clustering stops based on data quality rather than just depth limits
    \item Root level bias: More aggressive clustering at the root level with bonuses for additional clusters
    \item Representative selection: Uses highest-scoring images (most relevant to query) as cluster representatives
\end{itemize}

Core algorithm parameters:
\begin{itemize}
    \item Algorithm: Agglomerative clustering (Python scikit-learn implementation)
    \item Number of clusters: Dynamically determined (not fixed)
    \item Linkage: Ward linkage
    \item Affinity: Euclidean distance
\end{itemize}

Adaptive clustering parameters:
\begin{itemize}
    \item Max. hierarchy depth: 5 (with data-driven early stopping)
    \item Min. cluster size: 5
    \item Max. cluster size: 50
    \item Min. subcluster size: 40
    \item Min. silhouette: 0.1 (minimum silhouette score threshold)
\end{itemize}

Dynamic Cluster Count Parameters:
\begin{itemize}
    \item Root level minimum clusters: 3 (for initial split)
    \item Non-root level minimum clusters: 2
    \item Root level maximum clusters: 16
    \item Non-root level maximum clusters: 10
\end{itemize}

Quality Evaluation Parameters:
\begin{itemize}
    \item Distance metric for silhouette score: cosine
    \item Random state: 42
    \item Composite score weights:
    \begin{itemize}
        \item Silhouette score: Primary weight (1.0)
        \item Calinski-Harabasz index: Secondary weight (divided by 1000)
        \item Root level cluster bonus: 0.02 per additional cluster
        \item Root level penalty: 0.02 scaled by cluster ratio
        \item Non-root level penalty: 0.1 scaled by cluster ratio
    \end{itemize}
\end{itemize}

Stopping Criteria:
\begin{itemize}
    \item Minimum images for subclustering: 40
    \item Minimum images for meaningful split: Min. cluster size $\times$ 2
    \item Effective minimum size at root level: max(3, Min. cluster size // 2)
\end{itemize}

\section{Evaluation details}

\subsection{Counterbalance setting}
\label{app:counterbalance}

Table~\ref{tab:counterbalance} specifies for each participant of the user study which tasks (A$\sim$D) were solved with which tools, and their order. The Graeco-Latin square counterbalance consists of a set of 8 balanced sequences (4 rows + their reverses) that keeps positions perfectly balanced.

\begin{table}[ht]
 \footnotesize
 \caption{Counterbalance setting to assign the corresponding tool and task (A$\sim$D) pairs to each of the sixteen participants, in order from first to fourth.}
 \Description{List defining which tool was assigned to which participant in which order.}
 \label{tab:counterbalance}
 \begin{tabular}{ccccc}
   \toprule
   ID & First tool & Second tool & Third tool & Fourth tool \\
   \midrule
   1 & Clustering (B) & DiverXplorer (A) & Scroll (D) & Scroll+div (C) \\
   2 & Clustering (B) & DiverXplorer (A) & Scroll (D) & Scroll+div (C) \\
   3 & DiverXplorer (C) & Clustering (D) & Scroll+div (A) & Scroll (B) \\
   4 & DiverXplorer (C) & Clustering (D) & Scroll+div (A) & Scroll (B) \\
   5 & Scroll+div (D) & Scroll (C) & DiverXplorer (B) & Clustering (A) \\
   6 & Scroll+div (D) & Scroll (C) & DiverXplorer (B) & Clustering (A) \\
   7 & DiverXplorer (D) & Clustering (C) & Scroll+div (B) & Scroll (A) \\
   8 & DiverXplorer (D) & Clustering (C) & Scroll+div (B) & Scroll (A) \\
   9 & Scroll+div (C) & Scroll (D) & DiverXplorer (A) & Clustering (B) \\
   10 & Scroll+div (C) & Scroll (D) & DiverXplorer (A) & Clustering (B) \\
   11 & Scroll (B) & Scroll+div (A) & Clustering (D) & DiverXplorer (C) \\
   12 & Scroll (B) & Scroll+div (A) & Clustering (D) & DiverXplorer (C) \\
   13 & Clustering (A) & DiverXplorer (B) & Scroll (C) & Scroll+div (D) \\
   14 & Clustering (A) & DiverXplorer (B) & Scroll (C) & Scroll+div (D) \\
   15 & Scroll (A) & Scroll+div (B) & Clustering (C) & DiverXplorer (D) \\
   16 & Scroll (A) & Scroll+div (B) & Clustering (C) & DiverXplorer (D) \\   
   \bottomrule
\end{tabular}
\end{table}

\subsection{Participants}
\label{app:participants}

Table~\ref{tab:participants} summarizes the participants information of the user study (F/M stands for Female/Male).

\begin{table}[ht]
 \caption{Participants profile}
 \Description{List with the participant ID, gender, age range and years of experience of each designer than participated in our user study.}
 \label{tab:participants}
 \begin{tabular}{cccc}
   \toprule
   ID & Gender & Age range & Years of experience \\
   \midrule
   1 & F & 20s & 4 \\
   2 & M & 30s & 8 \\
   3 & F & 30s & 14 \\
   4 & F & 20s & 2 \\
   5 & F & 20s & 6 \\
   6 & F & 30s & 10 \\
   7 & F & 20s & 2 \\
   8 & F & 20s & 4 \\
   9 & F & 20s & 2 \\
   10 & F & 20s & 2 \\
   11 & F & 40s & 15 \\
   12 & M & 20s & 2 \\
   13 & M & 20s & 2 \\
   14 & F & 30s & 7 \\
   15 & F & 20s & 2 \\
   16 & M & 20s & 5 \\
 \bottomrule
\end{tabular}
\end{table}

\subsection{Questionnaire}
\label{app:questionnaire}

\begin{table*}[ht]
\centering
\caption{Questionnaire results summary (Q1$\sim$Q9). For each question: Mean and standard-deviation of each condition (i.e., exploration tool), Likert scale range, number of samples $N$, Friedman test $Q$, degrees of freedom $df$, and $p$ value.}
\Description{Table summarizing the answers of the questionnaire containing the question name, the mean and deviation for the scores assigned to each tool, the Likert scale, the number of samples/answers, que Friedman's test results, the degrees of freedom, and the $p$ value obtained}
\label{tab:questionnaire}
\begin{tabular}{lccccccccc}
\toprule
Question & \textit{Scroll} & \textit{Scroll+div} & \textit{Clustering} & \textit{DiverXplorer} & Likert & N & Q & df & $p$ \\ \midrule
Q1. Usefulness & 4.81 $\pm$ 1.17 & 4.75 $\pm$ 1.24 & 5.69 $\pm$ 1.01 & 5.81 $\pm$ 1.11 & 1$\sim$7 & 16 & 14.786 & 3 & $<.01$ \\
Q2. Smoothness & 4.62 $\pm$ 1.50 & 3.94 $\pm$ 1.34 & 5.25 $\pm$ 1.53 & 5.75 $\pm$ 1.24 & 1$\sim$7 & 16 & 15.287 & 3 & $<.001$ \\
Q3. Satisfaction & 4.75 $\pm$ 1.39 & 5.06 $\pm$ 1.12 & 5.69 $\pm$ 0.87 & 5.75 $\pm$ 1.29 & 1$\sim$7 & 16 & 8.193 & 3 & $<.05$ \\
Q4. Idea refinement & 4.69 $\pm$ 1.25 & 5.06 $\pm$ 1.12 & 5.56 $\pm$ 1.21 & 5.38 $\pm$ 1.45 & 1$\sim$7 & 16 & 6.432 & 3 & $n.s. (=0.092)$ \\
Q5. Data overall view & 5.06 $\pm$ 1.53 & 4.75 $\pm$ 1.44 & 5.88 $\pm$ 1.15 & 5.69 $\pm$ 1.35 & 1$\sim$7 & 16 & 8.944 & 3 & $<.05$ \\
Q6. Controllability & 4.12 $\pm$ 1.50 & 4.00 $\pm$ 1.41 & 5.56 $\pm$ 1.46 & 5.56 $\pm$ 1.41 & 1$\sim$7 & 16 & 16.370 & 3 & $<.001$ \\
Q7. Creativity support & 4.69 $\pm$ 1.45 & 5.38 $\pm$ 1.20 & 5.69 $\pm$ 1.08 & 5.62 $\pm$ 1.36 & 1$\sim$7 & 16 & 6.395 & 3 & $n.s. (=0.094)$ \\
Q8. Time saving & 4.81 $\pm$ 1.47 & 4.12 $\pm$ 1.82 & 5.44 $\pm$ 1.67 & 6.00 $\pm$ 1.10 & 1$\sim$7 & 16 & 12.508 & 3 & $<.01$ \\
Q9. Recommendation & 6.31 $\pm$ 2.06 & 5.81 $\pm$ 2.37 & 7.62 $\pm$ 1.89 & 7.94 $\pm$ 1.91 & 1$\sim$10 & 16 & 19.872 & 3 & $<.001$ \\
\bottomrule
\end{tabular}
\end{table*}

Intended to evaluate each tool individually, and based on recent creativity support tools~\cite{tao2025designweaver}, the questionnaire contained general questions regarding the usability and satisfaction of the user with the system.

\textit{Participant information.}
\begin{itemize}
    \item \textbf{Full name:} \underline{\hspace{2cm}}
    \item \textbf{Gender:} \underline{\hspace{2cm}}
    \item \textbf{Age range:} \underline{\hspace{2cm}}
    \item \textbf{Years of experience as a designer:} \underline{\hspace{2cm}}
\end{itemize}


\textit{Individual tool evaluation.}
\begin{itemize}
    \item \textbf{Rate these statements regarding the exploration system:} [1=Strongly disagree, 7=Strongly agree]
    \begin{itemize}
        \item Q1. It helped me succeed in the image exploration tasks.
        \item Q2. Reaching the desired image was easy/smooth.
        \item Q3. I am satisfied/confident with the images I reached.
        \item Q4. While using it, it helped me get a more refined idea of a design that would suit the task topic.
        \item Q5. It helped me get a general picture of the image data set.
        \item Q6. I felt in control of the exploration task.
        \item Q7. It supported my creativity.
        \item Q8. It saved me time in the exploration task.
    \end{itemize}
    \item \textbf{Recommendation score:} [1=Strongly disagree, 10=Strongly agree]
    \begin{itemize}
        \item Q9. I would recommend this tool to a friend or colleague to help with their creative work.
    \end{itemize}
    \item \textbf{Please let us know your impressions of the exploration tool:}
    \begin{itemize}
        \item Q10. What criteria did you use to select the final image?
        \item Q11. Specifically, what did you find intuitive/useful about the tool?
        \item Q12. Specifically, what did you find confusing/difficult about the tool? How would you improve it?
    \end{itemize}
\end{itemize}

Table~\ref{tab:questionnaire} summarizes the questionnaire scores for each question and tool. Since there are more than one degree of freedom (i.e., number tools minus 1), a Friedman($\chi^2$) test was conducted, and the $p$ values indicate that there were significant differences ($p<0.05$) among the exploration tools in all questions except Q4 and Q7.
Table~\ref{tab:likert_posthoc} summarizes the post-hoc analysis of the Friedman($\chi^2$) test for the questionnaire Likert scores. the Wilcoxon signed-rank test (Holm-adjusted $p_{\rm{Holm}}$ value) is used to analyze specific pairs of tools.

\begin{table*}
\centering
\caption{Post-hoc pairwise tests using the Wilcoxon signed-rank test (Holm-adjusted).}
\Description{List with the number of samples, Wilcoxon test and p_{Holm} comparing the scores of each pair of tools for each question of the questionnaire.}
\label{tab:likert_posthoc}
\begin{tabular}{l l c c c}
\toprule
Question & Pair & $n$ & $W$ & $p_{\text{Holm}}$ \\
\midrule
\multirow{6}{*}{Q1} & Clustering vs. DiverXplorer & \multirow{6}{*}{16} & 10.5 & $n.s.$ \\
 & Scroll vs. Clustering &  & 5.0 & $n.s. (=0.092)$ \\
 & Scroll vs. DiverXplorer &  & 18.0 & $n.s. (=0.092)$ \\
 & Scroll vs. Scroll+div &  & 25.5 & $n.s.$ \\
 & Scroll+div vs. Clustering &  & 11.0 & $<.05$ \\
 & Scroll+div vs. DiverXplorer &  & 4.5 & $<.05$ \\
\midrule
\multirow{6}{*}{Q2} & Clustering vs. DiverXplorer & \multirow{6}{*}{16} & 30.0 & $n.s.$ \\
 & Scroll vs. Clustering &  & 23.0 & $n.s.$ \\
 & Scroll vs. DiverXplorer &  & 36.0 & $n.s.$ \\
 & Scroll vs. Scroll+div &  & 22.5 & $n.s.$ \\
 & Scroll+div vs. Clustering &  & 17.0 & $n.s. (=0.062)$ \\
 & Scroll+div vs. DiverXplorer &  & 4.0 & $<.01$ \\
\midrule
\multirow{6}{*}{Q3} & Clustering vs. DiverXplorer & \multirow{6}{*}{16} & 23.0 & $n.s.$ \\
 & Scroll vs. Clustering &  & 5.0 & $n.s.$ \\
 & Scroll vs. DiverXplorer &  & 14.5 & $n.s.$ \\
 & Scroll vs. Scroll+div &  & 35.0 & $n.s.$ \\
 & Scroll+div vs. Clustering &  & 18.0 & $n.s.$ \\
 & Scroll+div vs. DiverXplorer &  & 18.0 & $n.s.$ \\
\midrule
\multirow{6}{*}{Q4} & Clustering vs. DiverXplorer & \multirow{6}{*}{16} & 33.0 & $n.s.$ \\
 & Scroll vs. Clustering &  & 26.0 & $n.s.$ \\
 & Scroll vs. DiverXplorer &  & 40.5 & $n.s.$ \\
 & Scroll vs. Scroll+div &  & 32.0 & $n.s.$ \\
 & Scroll+div vs. Clustering &  & 35.0 & $n.s.$ \\
 & Scroll+div vs. DiverXplorer &  & 33.0 & $n.s.$ \\
\midrule
\multirow{6}{*}{Q5} & Clustering vs. DiverXplorer & \multirow{6}{*}{16} & 20.0 & $n.s.$ \\
 & Scroll vs. Clustering &  & 20.5 & $n.s.$ \\
 & Scroll vs. DiverXplorer &  & 22.0 & $n.s.$ \\
 & Scroll vs. Scroll+div &  & 30.5 & $n.s.$ \\
 & Scroll+div vs. Clustering &  & 10.0 & $n.s.$ \\
 & Scroll+div vs. DiverXplorer &  & 14.0 & $n.s.$ \\
\midrule
\multirow{6}{*}{Q6} & Clustering vs. DiverXplorer & \multirow{6}{*}{16} & 10.0 & $n.s.$ \\
 & Scroll vs. Clustering &  & 4.0 & $n.s. (=0.086)$ \\
 & Scroll vs. DiverXplorer &  & 4.0 & $n.s. (=0.086)$ \\
 & Scroll vs. Scroll+div &  & 21.0 & $n.s.$ \\
 & Scroll+div vs. Clustering &  & 6.0 & $<.05$ \\
 & Scroll+div vs. DiverXplorer &  & 6.0 & $<.05$ \\
\midrule
\multirow{6}{*}{Q7} & Clustering vs. DiverXplorer & \multirow{6}{*}{16} & 27.0 & $n.s.$ \\
 & Scroll vs. Clustering &  & 9.0 & $n.s.$ \\
 & Scroll vs. DiverXplorer &  & 18.0 & $n.s.$ \\
 & Scroll vs. Scroll+div &  & 19.0 & $n.s.$ \\
 & Scroll+div vs. Clustering &  & 33.0 & $n.s.$ \\
 & Scroll+div vs. DiverXplorer &  & 33.0 & $n.s.$ \\
\midrule
\multirow{6}{*}{Q8} & Clustering vs. DiverXplorer & \multirow{6}{*}{16} & 24.0 & $n.s.$ \\
 & Scroll vs. Clustering &  & 8.0 & $n.s.$ \\
 & Scroll vs. DiverXplorer &  & 12.0 & $n.s. (=0.094)$ \\
 & Scroll vs. Scroll+div &  & 15.5 & $n.s.$ \\
 & Scroll+div vs. Clustering &  & 12.0 & $n.s. (=0.094)$ \\
 & Scroll+div vs. DiverXplorer &  & 2.0 & $<.05$ \\
\midrule
\multirow{6}{*}{Q9} & Clustering vs. DiverXplorer & \multirow{6}{*}{16} & 12.5 & $n.s.$ \\
 & Scroll vs. Clustering &  & 14.0 & $n.s.$ \\
 & Scroll vs. DiverXplorer &  & 12.0 & $n.s.$ \\
 & Scroll vs. Scroll+div &  & 17.5 & $n.s.$ \\
 & Scroll+div vs. Clustering &  & 12.0 & $n.s. (=0.057)$ \\
 & Scroll+div vs. DiverXplorer &  & 0.0 & $<.01$ \\
\bottomrule
\end{tabular}
\end{table*}

\end{document}